\documentclass[12pt]{article}
\usepackage{epsfig}
\usepackage{amsfonts}
\usepackage{amscd}
\usepackage{latexsym}
\usepackage{amsmath,amssymb}
\usepackage{verbatim}
\usepackage{setspace}
\usepackage{color}

\usepackage[textheight=9in, textwidth=6.5in, letterpaper]{geometry}
\def\half{{1\over 2}}
\numberwithin{equation}{section}

\def\ip{${\mathcal I}^+$}

\def\e{{\epsilon}}
\def\cs{{\cal S}}

\def\Psz{\Psi^{0} }
 \def\p{\partial}
 \def\bz{{\bar z}}
 
\def\0{{(0)}}
\def\1{{(1)}}
\def\2{{(2)}}
 \def\cL{{\cal L}}
\def\co{{\cal O}}

\def\ci{{\mathcal I}}

\def\<{\langle }
\def\>{\rangle }
\def\bw{{\bar w}}
\def\x{${\cal X}$}
\def\cM{{\cal M}}

\newcommand{\bea}{\begin{eqnarray}}
\newcommand{\eea}{\end{eqnarray}}
\newcommand{\be}{\begin{equation}}
\newcommand{\ee}{\end{equation}}
\newcommand{\ba}{\begin{align}}
\newcommand{\ea}{\end{align}}

\renewcommand{\epsilon}{\varepsilon}

   \makeatletter
  \let\over=\@@over \let\overwithdelims=\@@overwithdelims
  \let\atop=\@@atop \let\atopwithdelims=\@@atopwithdelims
  \let\above=\@@above \let\abovewithdelims=\@@abovewithdelims
\renewcommand\section{\@startsection {section}{1}{\z@}%
                                   {-3.5ex \@plus -1ex \@minus -.2ex}
                                   {2.3ex \@plus.2ex}%
                                   {\normalfont\large\bfseries}}

\renewcommand\subsection{\@startsection{subsection}{2}{\z@}%
                                     {-3.25ex\@plus -1ex \@minus -.2ex}%
                                     {1.5ex \@plus .2ex}%
                                     {\normalfont\bfseries}}

\begin{document}
\begin{titlepage}
\unitlength = 1mm
\ \\
\vskip 2cm
\begin{center}

{ \LARGE {\textsc{Semiclassical Virasoro Symmetry of the Quantum Gravity $\cs$-Matrix}}}

\vspace{0.8cm}
Daniel Kapec, Vyacheslav Lysov,  Sabrina Pasterski and Andrew Strominger

\vspace{1cm}

{\it  Center for the Fundamental Laws of Nature, Harvard University,\\
Cambridge, MA 02138, USA}

\begin{abstract}
It is shown that the tree-level $\cs$-matrix for quantum gravity in four-dimensional Minkowski space has a Virasoro symmetry which acts on the conformal sphere at null infinity. 
\end{abstract}

\vspace{1.0cm}

\end{center}

\end{titlepage}

\pagestyle{empty}
\pagestyle{plain}

\def\vx{{\vec x}}
\def\p{\partial}
\def\po{$\cal P_O$}

\pagenumbering{arabic}

\tableofcontents
\section{Introduction}

BMS$^+$ transformations \cite{bms} comprise a subset of diffeomorphisms which act nontrivially on future null infinity of asymptotically Minkowskian space times, or \ip.  BMS$^-$ transformations act isomorphically on past null infinity, or $\ci^-$.  A particular `diagonal' subgroup of the product group BMS$^+\times$BMS$^-$ has recently been shown \cite{asbms} to be a symmetry of gravitational scattering.  Ward identities of this diagonal symmetry relate $\cs$-matrix elements with and without soft gravitons. These $\cs$-matrix relations are not new \cite{hms}: they comprise Weinberg's soft graviton theorem \cite{steve}.  More generally, the connection to soft theorems provides a new perspective on asymptotic symmetries in Minkowski space \cite{as}. 

Over the decades a number of extensions/modifications  to the BMS group have been proposed: e.g. the Newman-Unti group \cite{nu}, the Spi group \cite{Ashtekar:1978zz} and the extended BMS group \cite{bt, Banks:2003vp}. A criterion is needed to decide whether or not such extensions are `physical'.  Here we adopt the 
pragmatic approach that a Minkowskian asymptotic symmetry is physical if and only if it provides nontrivial relations among $\cs$-matrix elements. We will view these $\cs$-matrix relations as a definition of the symmetry. 

{\sloppy
In this paper we will show that, at tree-level, quantum gravity in asymptotically Minkow\-skian  spaces in this sense has  a physical Virasoro symmetry. The symmetry is implied by a recently proven soft theorem \cite{fc} and acts 
(diagonally) on the conformal S$^2$ at $\ci$.
}

Our story begins with a conjecture of Barnich, Troessaert and Banks (BTB) \cite{bt, Banks:2003vp}.  BMS$^+$ has an $SL(2,C)$ Lorentz subgroup generated by the six global conformal Killing vectors (CKVs) on the S$^2$ at \ip.  Locally,  BTB showed that 
all of  the infinitely many CKVs preserve the same asymptotic structure at \ip\ and are hence also candidate asymptotic symmetry generators.  This larger set of vector fields  was a priori excluded in the original work of BMS, who demanded that they  be nonsingular everywhere on $S^2$. This restriction cuts the Virasoro group down to a mere $SL(2,C)$.  BTB conjectured that the true asymptotic symmetry group of \ip\ is the `extended BMS$^+$ group' generated by all CKVs. However it has not been clear if or in what sense the singular CKVs truly generate physical asymptotic symmetries.

Herein we consider, in the spirit of \cite{asbms}, a certain diagonal subgroup of \sloppy{(extended BMS$^+)$$\times$(extended BMS$^-$)}, denoted \x.  Ward identities are derived for a Virasoro subgroup of \x. They are found to involve a soft graviton insertion with the Weinberg pole projected out, leaving the finite subleading term in the soft expansion.  These Ward identities are in turn shown to be implied by a conjectured \cite{sac} soft relation schematically of the form
\be \label{suo} \lim_{\omega \to 0}\cM_{n+1}=S^{(1)}\cM_n.\ee
Here $\cM_{n+1}$ is an $n+1$-particle amplitude with a certain (pole-projected) energy $\omega$
soft graviton insertion, and  $S^{(1)}$ involves the soft graviton momentum as well as the energies and angular momenta of the  incoming and outgoing particles. Details are given below. The proof \cite{fc} of 
(\ref{suo}) for tree-level gravity amplitudes then implies a semiclassical Virasoro symmetry for the case of pure gravity.  This demonstrates that the singularities in the generic CKVs do not, at least in this context, prevent them from generating physical symmetries.\footnote{ It may alternatively be possible to reach this conclusion without appealing to direct computations such as in \cite{fc} by carefully regulating the singularities and analyzing their effects. We do not attempt such an analysis herein. } 

One might also hope to run the argument backwards and see to what extent the Virasoro symmetry of the $\cs$-matrix implies the soft relation (\ref{suo}). In the case of supertranslations, the argument can be run in both directions \cite{hms}. However here we encounter several obstacles, including 
the need for a prescription for handling the CKV singularities and some zero mode issues. We leave this to future investigations.\footnote{The Virasoro charges constructed in \cite{bt} may be useful for this purpose.}  Hence at this point the existence of a Virasoro symmetry is potentially a {\it weaker} condition than the validity of the soft relation (\ref{suo}).  

The analysis of \cite{asbms,hms} related two structures which have been well-established and thoroughly studied over the last half-century: 
  BMS symmetry and Weinberg's soft graviton theorem. Here the situation is rather different. We are relating two unestablished and understudied structures: asymptotic Virasoro 
symmetries and subleading soft graviton theorems. We hope the relation will illuminate both. In any case it is a rather different enterprise!

An important issue which we will not address is the quantum fate of the semiclassical Virasoro symmetry. 
Here the situation is currently up in the air. In \cite{Bern:2014oka, He:2014bga} it was shown that, in a standard regulator scheme, (\ref{suo}) receives IR divergent quantum corrections (at one loop only), which also make the $\cs$-matrix ill-defined in this scheme. However in \cite{Cachazo:2014dia},  the factor $S^{(1)}$ in  (\ref{suo}) relating the 5 and 4 point amplitude was found to  remain uncorrected at one loop in a scheme with the soft limit taken prior to removing the IR cutoff.\footnote{ \cite{Cachazo:2014dia} claims a result only for this one special case by direct computation. However, it has been suggested \cite{zbern} that, using \cite{Bern:1995ix}, a proof can be constructed in the scheme of \cite{Bern:2014oka, He:2014bga} that  all loop corrections to $S^1$ in (1.1) are linked to discontinuities arising from infrared singularities and hence  in the scheme of  \cite{Cachazo:2014dia} (with the soft limit  taken first) all loop corrections would disappear along with the discontinuities.} In the recent work \cite{sterman, Ware:2013zja} (see also \cite{white}) it was shown
that a properly defined $\cs$-matrix utilizing the gravity version of the Kulish-Faddeev construction \cite{Kulish:1970ut} is free of all IR divergences. This may be the proper context for the discussion, as it is hard to have a symmetry of an $\cs$-matrix without an $\cs$-matrix! Should it ultimately be found that (\ref{suo}) does receive scheme-independent corrections, one must then determine whether it implies a quantum anomaly in the asymptotic Virasoro symmetry (which is potentially weaker than (\ref{suo})), or a quantum deformation in its action on the amplitudes. Clearly highly relevant, but not yet fully incorporated into this discussion, is the low-energy theorem of Gross and Jackiw \cite{Gross:1968in} who use dispersion theory to show that there is no correction to the first three terms\footnote{$S^{(0)}$, $S^{(1)}$ and $S^{(2)}$ in the notation of \cite{Cachazo:2014dia}.} of the Born approximation to soft graviton-scalar scattering. This generalized the classic low-energy theorem for QED by Low \cite{Low:1954kd}. Progress on the gravity version was recently made by White \cite{white}.  Clearly, there is much to understand!

The existence of a Virasoro symmetry potentially has far-reaching implications for Mink\-owski quantum gravity in general.  However, at this point there are many basic unresolved points and it is too soon to tell  what or if they might be.  For example we do not know if the symmetry has quantum anomalies, what kind of representations appear,\footnote{They may not be the familiar ones from the study of unitary 2D CFT on the sphere.} the role of IR divergences or the connection to stringy Virasoro symmetries  \cite{as,Banks:2014iha}. Very recent developments indicate that these ideas, including the realization of the subleading soft theorem as a Virasoro symmetry, have a natural home in the twistor string \cite{Adamo:2014yya, twistwo}. Since the symmetry acts at the boundary, it is likely relevant to any holographic duality as long ago envisioned in \cite{ash}.

This paper is organized as follows. Section 2 establishes notation and reviews a few salient formulae for asymptotically flat geometries.  Section 3 describes the conjectured extended BMS$^\pm$  symmetry following \cite{bt}. In section 4 we define the diagonal subgroup \x\ of (extended BMS$^+)\times$(extended BMS$^-$) transformations, review Christodoulou-Klainerman (CK) spaces and define extended CK spaces by acting with \x.  A prescription is given to define  classical gravitational scattering from $\ci^-$ to \ip\  and shown to be symmetric under \x.  In section 5  the discussion of the quantum theory begins with the  the action of extended BMS$^\pm$ generators on in and out states.  A Ward identity is then derived which is equivalent to infinitesimal  \x -invariance of $\cs$. It relates amplitudes with and without a particular soft graviton insertion.  Finally in section 6 we give the detailed form of the soft relation (\ref{suo}) and show that it implies  the \x\ Ward identity. 
\section{Asymptotically flat geometry}
\subsection{Metrics}
A general asymptotically Minkowskian metric can be expanded in ${1 \over r}$ around \ip. In retarded Bondi coordinates it takes the form\footnote{We largely adopt  the notation of \cite{bt} to which we refer the reader for further details.}
\begin{align}
\label{eq:coord}
ds^2 =&-du^2  -2du dr +2r^2 \gamma_{z\bz} dzd\bz \notag  \\ 
&+\frac{2m_B}{r}du^2 + rC_{zz}dz^2 +r C_{\bz\bz} d\bz^2 +2g_{uz} dudz + 2g_{u\bz} dud\bz  +...,
\end{align}
where the first line is the flat Minkowski metric, $\gamma_{z\bz}$ ($D_z$) is the round metric (covariant derivative) on the unit $S^2$ and  
\be
g_{uz} = \frac12 D^z C_{zz} +\frac{1}{6r} C_{zz}D_z C^{zz} +\frac{2}{3r} N_z + \co(r^{-2}). 
\ee
The Bondi mass aspect $m_B$,  the angular momentum  aspect $N_z$ and $C_{zz}$ depend only on $(u, z, \bz)$ and not $r$. The outgoing news tensor  is defined by
\be
N_{zz}\equiv \p_u C_{zz}. 
\ee
\ip\ is the null surface $(r=\infty, u, z, \bz)$. We use the symbol $\ci^+_+$ ($\ci^+_-$)  to denote the future (past) boundary  of 
\ip\ at $(r=\infty, u=\infty, z, \bz)$ ($(r=\infty, u=-\infty, z, \bz)$). This is depicted in figure 1. 

There is an analogous construction on $\mathcal{I}^-$ with the metric given by
\begin{align}\label{sag}
ds^2&=-dv^2+2dvdr +2r^2\gamma_{z\bz} dzd\bar{z} \notag\\
&+\frac{2m_B^-}{r}dv^2 +rD_{zz}dz^2+rD_{\bz \bz}d\bz^2 +2g_{vz}dvdz+2g_{v\bz}dvd\bz +...,
\end{align}
with 
\be
g_{v z}= -\frac12 D^z D_{zz} -\frac{1}{6r} D_{zz}D_z D^{zz} - \frac{2}{3r} N^-_z + \co(r^{-2}). \ee
The $\ci^-$ coordinate $z$ in (\ref{sag}) is antipodally related to the \ip\ coordinate $z$ in (\ref{eq:coord}) in the sense that, for flat Minkowski space, a null geodesic begins and ends at the same value of $z$. Put another way, in the conformal compactifcation of asymptotically flat spaces, all of $\ci$ is generated by null geodesics which run through spatial infinity $i^0$. These generators have the same constant $z$ value on both \ip\ and $\ci^-$. 
The  incoming news tensor is defined by
\be
M_{zz}\equiv \p_v D_{zz}.
\ee

\begin{figure}
\centering
\includegraphics[width=1.0\textwidth]{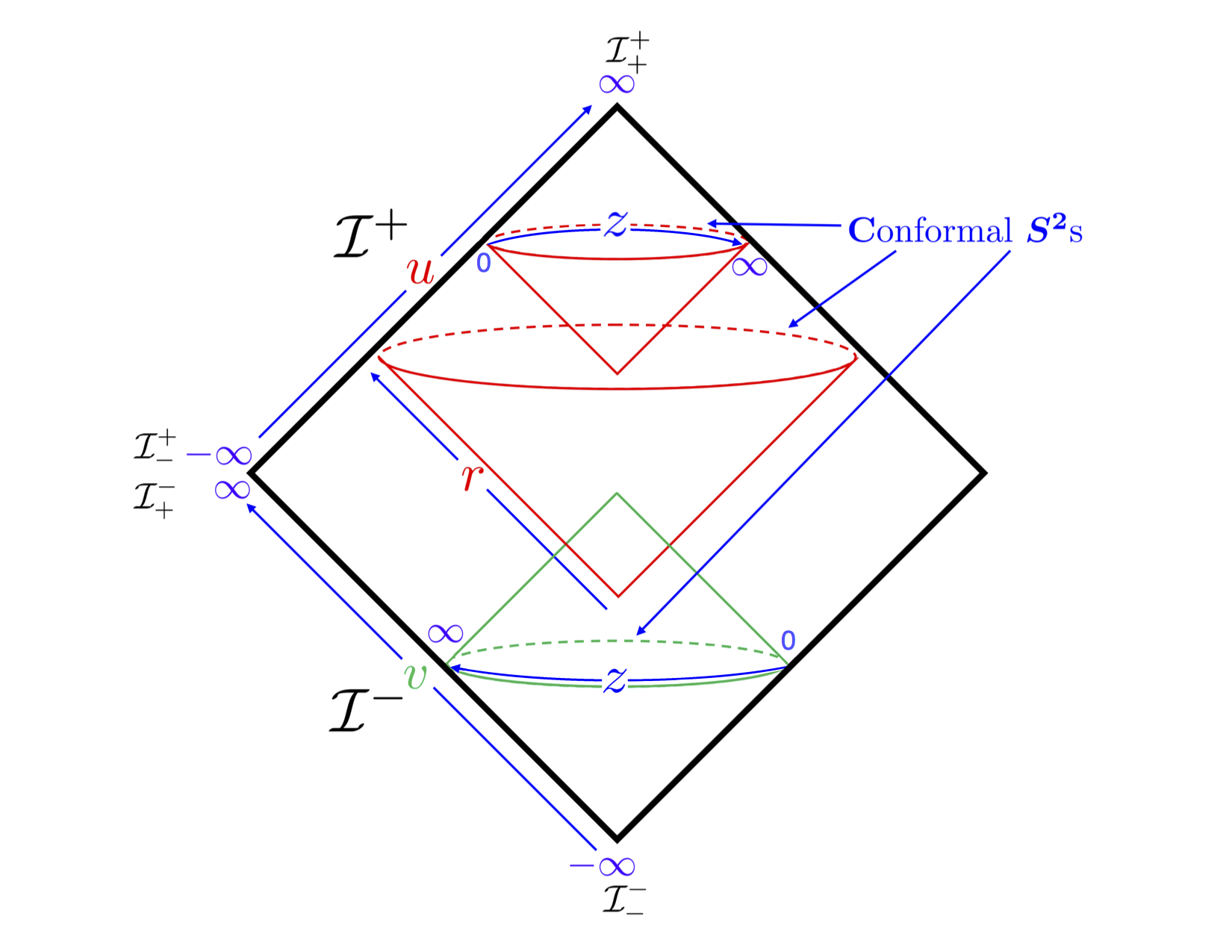}
 \caption{Penrose diagram for Minkowski space.  Near \ip\ surfaces of constant retarded time $u$ (red) are cone-like and intersect \ip\ in a conformal $S^2$ parametrized by $(z,\bz)$.   Cone-like surfaces of constant advanced  time $v$ (green) intersect $\ci^-$ in a conformal $S^2$ also parametrized by $(z,\bz)$. The future (past) $S^2$ boundary of \ip\ is labelled $\ci^+_+$ ($\ci^+_-$), while the future (past) boundary of  $\ci^-$ is labelled $\ci^-_+$ ($\ci^-_-$).}
\end{figure}

When expanding about  flat Minkowski space we sometimes  employ flat coordinates in which the flat metric takes the form
\be ds_F^2=\eta_{\mu\nu}dx^\mu dx^\nu.\ee
These are related to Bondi coordinates in flat space by
\bea\label{flt}
x^0&=&u+r=v-r,  \cr
  x^1+ix^2&=&{2rz \over 1+z\bz},\cr
  x^3&=& {r(1-z\bz) \over 1+z\bz}.         
  \eea

\subsection{Constraints}
The data in (\ref{eq:coord}) are related by the constraint equations $G_{\mu\nu}=\mathcal{T}^M_{\mu\nu}$, where $\mathcal{T}^M_{\mu\nu}$ is the matter stress tensor and we adopt units in which $8\pi G=1$. The leading term in the expansion of the $G_{uu}$ constraint equation about \ip\ is 
\be
\p_u m_B=\frac{1}{4}D_z^2 N^{zz}+\frac{1}{4}D_{\bz}^2N^{\bz\bz}  -\half T^M_{uu}- \frac{1}{4}N_{zz}N^{zz}, 
\ee
where 
\be \label{mt}
T^M_{\mu\nu}(u, z,\bz)=\lim\limits_{r\to \infty}  r^2\mathcal{T}^M_{\mu\nu }(u, r, z,\bz)  \ee
is the rescaled matter stress tensor which we have assumed falls of like ${1 \over r^2}$ near \ip. 
The $G_{uz}$ constraint gives
\begin{align}
\p_u N_z &= -\frac14 ( D_{z}D_{\bz}^2 C^{\bz\bz}- D_{z}^3 C^{zz})- {T}^M_{uz}+\p_z m_B+\frac{1}{16} D_z \p_u(C_{zz}C^{zz}) \\
&~~~~-\frac14  N^{zz}D_z C_{zz} -\frac14  N_{zz}D_z C^{zz} 
-\frac14 D_z ( C^{zz} N_{zz} - N^{zz}C_{zz}) . \notag
\end{align}
Given the Bondi news, $m_B$, $N_z$ and $C_{zz}$ are all determined up to $u$-independent integration constants which are discussed below.  The $\ci^-$ constraints are \be
\p_v m_B^-=\frac{1}{4}D^2_zM^{zz}+\frac{1}{4}D_{\bz}^2M^{\bz\bz} +\frac{1}{2}{T}^M_{vv}+ \frac{1}{4}M^{zz}M_{zz} ,
\ee
\begin{align}
\p_v N^-_z& = \frac14 (D_zD_{\bz}^2 D^{\bz\bz} - D^3_z  D^{zz}) -{T}^M_{vz}-\p_zm^-_B+\frac{1}{16} D_z \p_v(D_{zz}D^{zz}) \notag \\
&~~~~-\frac14  M^{zz}D_z D_{zz} -\frac14  M_{zz}D_z D^{zz} 
-\frac14 D_z ( D^{zz} M_{zz} - M^{zz}D_{zz}) .
\end{align}

\section{ Extended BMS$^\pm$ transformations }

The extended BMS$^+$  group has been  proposed \cite{bt,Banks:2003vp} as the asymptotic symmetry group at \ip\ of gravity on asymptotically flat spacetimes.  It is generated by  vector fields $\xi^+$ that  locally preserve the asymptotic form (\ref{eq:coord}) of the metric  at \ip 
\begin{align}
\cL_{\xi^+} g_{ur} &=\co(r^{-2}),\;\;\;\;\cL_{\xi^+} g_{uz} =\co(1),\;\;\;\cL_{\xi^+} g_{zz} =\co(r) ,\;\;\;
\cL_{\xi^+} g_{uu} =\co(r^{-1}). 
\end{align}
All such vector fields near \ip\ are of the form 
\begin{align}\label{vf}
\xi^+  =&(1+\frac{u}{2r}) Y^{+z}\p_z - \frac{u}{2r}  D^\bz D_z Y^{+z} \p_{\bz} 
-\frac12 (u+r  )D_z  Y^{+z} \p_r+{ u \over 2} D_z Y^{+z} \p_u + c.c.\\ \notag
&+f^+\p_u  - \frac1r (D^zf^+\p_z + D^{\bz}f^+\p_{\bz}) +D^zD_zf^+\p_r ,
\end{align}
where  $f^+$ is an arbitrary function on $S^2$ and here and elsewhere we suppress  (in some cases metric-dependent) terms which are further subleading in ${1 \over r}$ and irrelevant to our analysis: see \cite{bt} for a recent treatment specifying these terms. 
$Y^+$ must be a conformal Killing vector on $S^2$ which obeys the equation
\be \label{dsx} \p_\bz Y^{+z}=0. \ee
Globally there are six real vectors fields in an antisymmetric matrix $Y^z_{\mu\nu}$  obeying (\ref{dsx}): 
\be\label{alr}
\begin{array}{c}
 Y^z_{12} =iz,\;\;  Y^z_{13} = -\frac 12 (1+z^2) ,\;\; Y^z_{23} = -\frac i2 (1-z^2), \\
Y^z_{03} = z,\;\; Y^z_{01} =-\frac 12 (1-z^2),\;\; Y^z_{02} =-\frac i 2 (1+z^2).
\end{array}
\ee These generate the six  Lorentz boosts and rotations on \ip.  Locally there are infinitely many solutions of the form $Y^z\sim z^n$ with poles somewhere on the sphere. In their original work \cite{bms},  
BMS excluded these singular vector fields. However in this paper we shall explore the conjecture of \cite{bt,Banks:2003vp} that all of these `superrotations' should be included as part of the asymptotic symmetry group. 

The extended BMS$^+$  group is a semi-direct product of  superrotations with supertranslations. The supertranslations were recently analyzed in \cite {asbms,hms}.
For notational brevity we henceforth  consider only the  superrotation subgroup which has $f^+=0$ in (\ref{vf}) and reduces to
\be \label{xip}
\xi^+  =(1+\frac{u}{2r}) Y^{+z}\p_z - \frac{u}{2r} D^{\bz} D_z Y^{+z} \p_{\bz} 
-\frac12 (u+r  )D_z  Y^{+z} \p_r+{ u \over 2} D_z Y^{+z} \p_u + c.c.
\ee
This maps \ip\ to itself via 
\be \xi^+|_{\ci^+}=  Y^{+z}\p_z +{ u \over 2} D_z Y^{+z} \p_u + c.c.\ee
Similarly on $\mathcal{I}^-$ we have  BMS$^-$ vector fields parametrized by $Y^-$
\be \label{xim}
\xi^-  =(1-\frac{v}{2r}) Y^{-z}\p_z +\frac{v}{2r}D^{\bz}D_z Y^{-z} \p_{\bz} 
-\frac12 (r-v  )D_z  Y^{-z} \p_r+\frac v2  D_z Y^{-z} \p_v + c.c. 
\ee
Infinitesimal BMS$^+$ transformations act on the Bondi-gauge metric components as
\begin{align}
\delta_{Y^+} C_{zz}& =  { u \over 2}( D_z Y^{+z} +D_{\bz} Y^{+\bz}) \p_u C_{zz}+\cL_{Y^+} C_{zz} -\frac12 ( D_z Y^{+z} +D_{\bz} Y^{+\bz})  C_{zz} - u D_{z}^3 Y^{+z}, \notag \\ 
\delta_{Y^+} N_{zz} &\equiv \p_u \delta C_{zz}  =   { u \over 2} ( D_z Y^{+z} +D_{\bz} Y^{+\bz})  \p_u N_{zz}+\cL_{Y^+} N_{zz} -  D_{z}^3 Y^{+z}.  \end{align}
Similarly at $\mathcal{I}^-$ 
\begin{align}
\delta_{Y^-} D_{zz}& =  \frac v2 ( D_z Y^{-z} +D_{\bz} Y^{-\bz})  \p_v D_{zz}+\cL_{Y^-} D_{zz} -\frac12 ( D_z Y^{-z} +D_{\bz} Y^{-\bz}) D_{zz} +v D_{z}^3 Y^{-z},\notag  \\ 
\delta_{Y^-} M_{zz} &=   \frac v 2 ( D_z Y^{-z} +D_{\bz} Y^{-\bz})  \p_v M_{zz}+\cL_{Y^-} M_{zz} +  D_{z}^3 Y^{-z}.  
\end{align}

\section{\x\  transformations}

BMS$^\pm$ symmetries act on the physical data at $\ci^\pm$ while preserving certain asymptotic structures  such as the symplectic form \cite{ash}.  They are not  themselves symmetries of gravitational scattering: that is given some solution $(D_{zz}, C_{zz})$ of the gravitational scattering problem we cannot get a new one by acting with an element of BMS$^+$ or BMS$^-$. However for the case of supertranslations, it was argued in \cite{asbms} that  a certain diagonal subgroup of BMS$^+\times $BMS$^-$ is a symmetry of gravitational scattering in a suitable neighborhood \cite{ck} of flat space.  This subgroup is generated by pairs of $SL(2,C)$ Killing vector fields and supertranslations $(Y^+,f^+;Y^-,f^-)$ obeying 
\be \label{sv} Y^{+z}(z)=Y^{-z} (z)\equiv Y^z(z),~~f^+(z,\bz)=f^-(z,\bz)\equiv f(z,\bz),\ee
with the understanding that the coordinate $z$ is constant along null generators of $\ci$ as they pass from $\ci^-$ to \ip\ through spatial infinity $i^0$ in the conformal compactification of the spacetime. This means that points labelled by the same value of $z$ on $\ci^-$ and \ip\ lie at antipodal angles form the origin.  This antipodal identification may sound a little odd at first, but in fact is required in order for the subgroup (\ref{sv}) to contain the usual global Poincare transformations. 

In this paper we are interested in {\it extended} BMS$^+\times $BMS$^-$ transformations. We denote by \x\ the subgroup of these transformations generated by vector fields asymptotic to $(\xi^+,\xi^-)$ on $(\ci^+,\ci^-)$ subject to (\ref{sv}), where now $Y^z$ is any of the infinitely many conformal Killing vectors on the sphere. Elements of \x\  transform a solution $(D_{zz}, C_{zz})$ of the gravitational scattering problem to a new one   $(D'_{zz}, C'_{zz})$ with different final and initial data. We will argue below that the new data is a new solution of the scattering problem.

\subsection{Christodoulou-Klainerman spaces }
	 We are interested in asymptotically flat solutions of the Einstein equation which revert to the vacuum in the  far past and future. In particular we want to remain below the threshold for black hole formation.  We will adopt the rigorous definition of such spaces given by Christodoulou and Klainerman (CK) \cite{ck}  who also proved their global existence and  analyzed their asymptotic behavior. 

CK  studied asymptotically flat initial data in the center-of-mass frame on a maximal spacelike 
slice for which the Bach tensor $\e^{ijk}D^ {(3)}_jG^{(3)}_{kl}$ of the induced three-metric decays like $r^{-7/2}$ (or faster) at spatial infinity 
and the extrinsic curvature like $r^{-5/2}$. This implies that in normal coordinates about infinity the  leading part of the three-metric has the (conformally flat) Schwarzschild form, with corrections which decay like $r^{-3/2}$. CK showed that all such initial data which moreover 
satisfy a global smallness condition give rise to a global, i.e. geodesically complete, solution.  We will refer to these solutions as CK spaces. 

The smallness condition is satisfied in a finite neighborhood of Minkowski space, so this result established the stability of Minkowski space. Moreover many asymptotic properties of CK spaces at null infinity were derived in detail, see \cite{Christodoulou:1991cr} for a summary. Here we note that  the Bondi news $N_{zz}$   vanishes on the boundaries of \ip\ as \be \label{ff} N_{zz}(u) \sim |u|^{-3/2},\ee
or faster. Similarly on $\ci^-$ \be M_{zz} (v) \sim |v|^{-3/2} \ee
or faster.
The Weyl curvature component $\Psz_2$ which in coordinates (\ref{eq:coord}) is given by \bea\Psz_2(u,z,\bz)&\equiv& -\lim_{r \to \infty}({r}C_{u\bz r z}\gamma^{z\bz})
 \cr &=&-m_B-{1 \over 4} C_{zz}N^{zz}+\frac14 (D^zD^z C_{zz}- D^{\bz}D^{\bz}C_{\bz\bz})\eea obeys  \be \label{io}\Psz_2 |_{\ci^+_+}=0,\ee
while at $u=-\infty$
\be \label{iob}\Psz_2 |_{\ci^+_-}=-GM,\ee where $G$ is  Newton's constant and $M$  is the ADM mass. Similar results pertain to  $\ci^-$.

In this paper we consider generalizations of  pure gravity which include coupling  massless matter which dissipates at late (early) times on \ip\ ($\ci^-$) so that the system begins and ends in the vacuum.  The CK analysis has not been fully generalized to this case, although there is no obvious reason analogs of (\ref{ff})-(\ref{iob}) might not still pertain to a suitably defined neighborhood of the gravity+matter vacuum. In the absence of such a derivation (\ref{ff})-(\ref{iob})  will simply be imposed,  in the matter-coupled  case,  as restrictions on the solutions under consideration. 

\subsection{Classical  gravitational scattering}

 The classical problem of gravitational scattering is to find the outgoing data at \ip\ resulting from the evolution of given  data 
on $\ci^-$. We take the incoming data to be $D_{zz}(v,z,\bz)$ and the outgoing data to be $C_{zz}(u,z,\bz)$. The remaining metric components on $\ci$ are then determined by constraints. We consider the geometries in the neighborhood of flat space defined by CK, which have $m_B=0$  ($m^-_B=0$) at $\ci^+_+$    ($\ci^-_-$).  In particular we remain below the threshold for black hole formation. 

A CK geometry, as described in $(t,r,\theta, \phi)$  coordinates, does not quite provide a solution to this scattering problem. To find the in (out) data, one must perform a coordinate transformation to ingoing (outgoing) Bondi coordinates and determine $D_{zz}$ 
($C_{zz}$). This procedure is not unique:  the coordinate transformations are ambiguous up to extended BMS$^\pm$  transformations on \ip\ or $\ci^-$. $D_{zz}$ 
 and $C_{zz}$ are not invariant under these transformations.  Hence a  solution of the scattering problem requires a prescription for fixing this ambiguity. 
A prescription to fix this ambiguity is to demand that 
\bea \label{zro} D_{zz}|_{\ci^-_+}=C_{zz}|_{\ci^+_-}=0.
\eea
It was shown in \cite{asbms} that the falloffs (\ref{ff})-(\ref{iob}) imply this is always possible. 
One may then integrate the constraint equations to determine $D_{zz}$ 
and $C_{zz}$, which will not in general vanish at $\ci^+_+$ and $\ci^-_-$.

This prescription does not give all near-flat solutions of the scattering problem. Indeed,  all such solutions are in the center-of mass frame and have vanishing ADM three-momentum. However, given any solution of the scattering problem obeying (\ref{zro}), a new one with nonzero three-momentum may be obtained simply by acting with the boost element of \x . More generally, our prescription to define gravitational scattering is to take all solutions obtained by doing arbitrary \x\ transformations on the solutions obeying (\ref{zro}). We shall refer to such scattering geometries, complete with $\ci^\pm$ data,  as { \it extended} CK spaces. Acting with an arbitrary finite conformal transformation $w(z)$ followed by an arbitrary finite supertranslation $f$ on (\ref{zro}) leads to the asymptotic behaviors for large negative $u$ and positive  $v$\footnote{ Interestingly the news tensor at the boundary $\ci^+_-$ obeys $ N_{ww}|_{\ci^+_-}=-2({\p_wz})^{1/2} \p_w^2 ({{\p_wz})^{-1/2}},$
which is the transformation law for a 2D CFT stress tensor.  }
\bea 
\label{eq:asym}
C_{ww}(u,w,\bw)& \sim  -2 u   ({\p_wz})^{1/2} \p_w^2 ({{\p_wz})^{-1/2}}-2D_w^2f +\co(u^{-3/2}),\cr
D_{ww}(v,w,\bw)& \sim 2v  ({\p_wz})^{1/2} \p_w^2 ({\p_wz})^{-1/2} +2D_w^2f +\co(v^{-3/2}). \eea
We also have the relations at all the boundaries of $\ci^\pm$
\bea \p_\bz N_{zz}|_{\ci^+_\pm}&=0,\cr  \p_\bz M_{zz}|_{\ci^-_\pm}&=0,\cr
[ D_\bz^2 C_{zz}-D_z^2 C_{\bz\bz}]_{\ci^+_\pm}&=0,\cr  [ D_\bz^2 D_{zz}-D_z^2 D_{\bz\bz}]_{\ci^-_\pm}&=0. \eea
\section{ \x\ Ward identity }
\subsection{Quantum states}

In the quantum theory,  incoming (outgoing) states on $\ci^-$ (\ip) are presumed to form representations of extended BMS$^-$ (BMS$^+$).  In this subsection we will describe the action of an infinitesimal  Virasoro transformation $\delta_{Y}$ parameterized by $Y^z$ on a generic Fock-basis in-state. For $\ci^-$ we define \be Q^-(Y^-)|in \rangle= -i\delta_{Y^-}|in \rangle,\ee
and similarly we define $Q^+(Y^+)$ on \ip.\footnote{ Explicit expressions for the proper BMS$^\pm$ charges as integrals of fields on $\ci$ were worked out in detail in \cite{hms} and shown to generate the proper BMS$^\pm$ symmetries.  Expressions for the Virasoro charges $Q^\pm$ are given in \cite{bt}, but were not shown to generate the symmetries.  In this paper such explicit expressions will not be needed: transformation laws for the states suffice.} $Q^-$ may be decomposed into a hard and soft part as 
\be Q^-=Q^-_H+Q^-_S, \ee
where $Q^-_H$ generates the diffeomorphism $\xi^-(Y^-)$ on the incoming hard particles, 
and $Q^-_S$ creates a soft graviton. Let us denote  an in-state comprised of $n$ particles with energies $E_k$  incoming at points $ z_k$ for $k=1,\dots ,n$ on the conformal $S^2$ by  
\be |z_1,z_2,...\>. \ee  
Then the hard action is simply to act with $\xi^{-\mu}\p_{k\mu}$ on each scalar particle 
\be \label{oj}  Q^-_H|z_1,z_2,...\>=-i\sum_k \left( Y^{-z}(z_k)\p_{z_k}-\frac {E_k}{2} D_zY^{-z}(z_k) \p_{E_k}\right) |z_1,z_2,...\>, \ee
Here $-(1+E_k \p_{E_k})$ arises from the Fourier transform of $v\p_v$, and the coefficient of $D_zY^{-z}$ is shifted by one half as in \cite{as} due to the $r\p_r$ term in (\ref{xim}). For spinning particles we must replace $Y^{-z}(z_k)\p_{z_k}$ with the Lie derivative $\cL_{Y^-(z_k)}$.\footnote{\label{hlie}More explicitly if we have a particle of helicity $h$, and Rindler energy $-iv\p_v=E_R$, the parentheses in (\ref{oj}) are of the form
$ Y^{z}\p_{z}+Y^{\bz}\p_{\bz}+h_R D_zY^{z}+h_L D_\bz Y^{\bz}$ where for helicity $h$, 
the `conformal weights' (see e.g. \cite{as}) are $h_R = \frac{h}{2}-\frac12 E\p_E =\frac12 ( h +1+iE_R) ,~h_L =-\frac{h}{2} -\frac12 E\p_E   = \frac12( -h +1+iE_R )$.}

 To determine $Q^-_S$, note that the inhomogeneous transformation of the incoming Bondi news $M_{zz}$ is 
\be \delta_{Y^-}M_{zz}(v,z,\bz)= D_z^3Y^{-z}.\ee
The action of $Q^-_S$ on a state must implement this shift. It follows that 
\be \label{ate} [Q^-_{S}, M_{zz}] =  -iD_z^3 Y^{-z}. \ee
Using the commutator \cite{ash}
\be [M_{\bz\bz}(v,z,\bz),M_{ww}(v',w,\bw)]=2i \gamma_{z\bar{z}}\delta^{2}(z-w)\p_v\delta(v-v'), \ee
one concludes that, up to a total derivative commuting with $M_{zz}$,
\be
Q^-_{S} =  \frac{1}{2} \int_{\mathcal{I}^-} dv  d^2z 
  D_z^3  Y^{-z} vM^z_{~\bz}. \ee
  This reproduces the linear term in the full expression for the charge given in \cite{bt}.\footnote{The formula in \cite{bt} differs by a total derivative which improves the large $|v|$ behavior and may be essential in a more general context. The slightly simpler expression here is sufficient for the present purpose.}
  $Q^-_S$ is a zero-frequency operator (because of the $v$ integral) linear in the metric fluctuation. 
 Acting on the in - vacuum, it creates a soft graviton with polarization tensor proportional to $D_z^3Y^{-z}$. The explicit form of the momentum space creation operator will be constructed below in subsection 5.3.  Altogether then $Q^-$ maps the $n$-particle states into themselves plus an $n$-hard+1-soft state: 
\be Q^-|z_1,z_2,...\>=-i\sum_{k =1}^n \left( Y^{-z}(z_k)\p_{z_k}-\frac{E_k}{2} D_zY^{-z}(z_k) \p_{E_k}\right) |z_1,z_2,...\> ~ +Q^-_S|z_1,z_2,...\> . \ee
Similarly Virasoro  transformations on \ip\ are decomposed as 
\be Q^+=Q^+_H+Q^+_S\ee
and we denote  out-states comprised of $m$ particles with energies $E_k$  outgoing  at points $ z_k$ for $k=n+1,...n+m$ by  
\be \<z_{n+1},z_{n+2},...|. \ee
One finds
\be
\begin{array}{ll}
& \<z_{n+1},z_{n+2},...|Q^+=i\sum\limits_{k=n+1}^{n+m} \left( Y^{+z}(z_k)\p_{z_k}-\frac{E_k}{2} D_zY^{+z}(z_k) \p_{E_k}\right)\<z_{n+1},z_{n+2},...| ~~~ \\
 &~~~~~~~~~~~~~~~~~~~~~~~~~~~~~~~~~~~~~~~~~~~~~~~~~~~~~~~~~~~~~~~~~~~~~~~~~~~~+\<z_{n+1},z_{n+2},...|Q^+_S , 
 \end{array}
 \ee
where 
\be Q^+_{S} = -\frac{1}{2} \int_{\mathcal{I}^+} du d^2z
  D_z^3  Y^{+z} uN^z_{~\bz}.\ee

\subsection{\x-invariance of $\cs$}
In this section we derive a quantum Ward identity from the assumption that \x-invariance survives quantization. The quantum version of  infinitesimal \x\ invariance of classical gravitational scattering is, using  (\ref{sv}) \be\label{dst}  \<out| Q^+(Y)\cs-\cs Q^-(Y)|in \>=0,\ee
for any pair of in and out states ($|in\>, |out\>$).  Let us define the normal-ordered soft graviton insertion
\be  :Q_S(Y)\mathcal{S}:=  Q_{S}^+ (Y)\mathcal{S} - \mathcal{S} Q^-_{S}(Y). \ee 
(\ref{dst}) is then the Ward identity
\be\label{conj}
\begin{array}{ll}
&\<z_{n+1},z_{n+2},... |:Q_S \mathcal{S}:|z_1,z_2,...\>=\\
&~~~~~~~~~~~~~~ -i \sum\limits_{k=1}^{n+m} \left( Y^z(z_k)\p_{z_k}-\frac{E_k}{2}D_zY^z(z_k) \p_{E_k}\right) \<z_{n+1},z_{n+2},...|\mathcal{S} |z_1,z_2,...\>,
\end{array}
\ee
where the $k$ sum now runs over both in and out  particles and again for spinning particles the Lie derivative replaces the ordinary one on the right hand side. This relates the derivatives of any $\cs$-matrix element to the same $\cs$-matrix element with a particular soft graviton insertion. 
\subsection{Mode expansions}
We wish to express $Q^\pm_{S}$   in terms of standard momentum space  soft graviton creation and annihilation operators. 
The flat space graviton mode expansion  is\footnote{Here we take $g_{\mu\nu}=\eta_{\mu\nu}+\sqrt{32\pi G}h_{\mu\nu}= \eta_{\mu\nu}+2h_{\mu\nu}$.}
\begin{equation}
h^{\mathrm{out}}_{\mu\nu}(x) = \sum\limits_{\alpha=\pm} \int \frac{ d^3q}{(2\pi)^3} \frac{1}{2 \omega_q} \left[ \e^{\alpha*}_{\mu\nu} ({ \vec q})a^{\mathrm{out}}_\alpha ({\vec q}) e^{i q \cdot x} + \e^\alpha_{\mu\nu}({\vec q}) a^{\mathrm{out}}_\alpha ({\vec q})^\dagger   e^{- i q \cdot x} \right],
\end{equation}
where $q^0 = \omega_q = | {\vec q}|$, $\alpha=\pm$ are the two helicities and
\be\label{rrd}
[a^{\mathrm{out}}_\alpha ({\vec q}), a^{\mathrm{out}}_\beta ({\vec{q'}})^\dagger ]= 2\omega_q\delta_{\alpha\beta}(2\pi)^3\delta^3 \left( {\vec q} - {\vec q}' \right).
\ee
The outgoing gravitons with momentum $q$ correspond to final-state insertions of $a^{\mathrm{out}}_\alpha ({\vec q})$. It is convenient to parametrize the graviton four-momentum
by  $(\omega_q,w,\bw)$ \begin{equation}\label{gravmom}
q^\mu = \frac{\omega_q}{1 + w {\bar w}} \left( 1 + w {\bar w} , w + {\bar w} ,  i \left( \bw -  w\right), 1 - w {\bar w}  \right),
\end{equation}
with  polarization tensors 
\bea 
\e^{\pm\mu\nu} &=&\e^{\pm\mu}\e^{\pm\nu}, \cr
{ \e}^{+\mu}( {\vec q} ) &=& \frac{1}{\sqrt{2}} \left( {\bar w}, 1, - i, - {\bar w} \right),  \cr
{\e}^{-\mu}({\vec q} ) &=& \frac{1}{\sqrt{2}} \left( w , 1,   i, - w  \right).
\eea
These obey  $\e^{\pm\mu}q_\mu= \e^{\pm\mu}{}_\mu=0$ and 
\be \label{vv}\e_\bz^+ \left({\vec q} \right) = \p_\bz x^\mu\e^+_\mu \left( \vec{q} \right)= \frac{ \sqrt{2} r \left( 1 + z\bar{w} \right)}{ \left( 1 + z {\bar z} \right)^2 }   ,~~~~\e_\bz^- \left({\vec q} \right) = \p_\bz x^\mu\e^-_\mu \left( \vec{q} \right)  = \frac{ \sqrt{2} r { z} \left( { w} - { z} \right) }{ \left( 1 + z {\bar z} \right)^2 }.\ee
In retarded Bondi coordinates 
 \begin{equation}
C_{\bz\bz}(u,z,\bz) = 2 \lim_{r\to\infty} \frac{1}{r} h^{\mathrm{out}}_{\bz\bz}(r,u,z,\bz).
\end{equation}
 Using $h^{\mathrm{out}}_{\bz\bz} = \p_\bz x^\mu \p_\bz x^\nu h^{\mathrm{out}}_{\mu\nu}$ and the mode expansion 
\be
\label{eq:saddle}
C_{\bz\bz} = 2 \lim\limits_{r\to\infty} \frac{1}{r} \p_\bz x^\mu \p_\bz x^\nu   \sum\limits_{\alpha=\pm} \int \frac{ d^3q}{(2\pi)^3} \frac{1}{2 \omega_q} \left[ \e^{\alpha*}_{\mu\nu} ({ \vec q})a^{\mathrm{out}}_\alpha ({\vec q}) e^{- i \omega_q u - i \omega_q r \left( 1 - \cos\theta \right) } + h.c. \right],
\ee
where $\theta$ is the angle between $\vec{x}$ and $\vec{q}$.  {This integral is dominated for large $r$ by the contribution near $\theta$=0:}
\be
\label{eq:czz}
C_{\bz\bz}=-\frac{i}{4\pi^2}\hat{\e}_{\bz\bz}^{+}\int_0^\infty d\omega_q[a^{\mathrm{out}}_-(\omega_q\hat{x})e^{-i\omega_qu}-a^{\mathrm{out}}_+(\omega_q\hat{x})^\dagger e^{i\omega_qu}].
\ee
 Here, $\hat{x}$ is parameterized by $(z,\bar{z})$
\be
\hat{x} \equiv \frac{\vec{x}}{r} = \frac{1}{1+z\bz}(z+\bz, i(\bz-z), 1-z\bz)
\ee
 and 
\be
\hat{\e}^{+ }_{\bz\bz} = \frac{\p_\bz x^\mu \p_\bz x^\nu }{r^2} \e_{\mu\nu}^{+}  = \frac{2}{(1+z\bz)^2}.
\ee
Define:
\be
N^{\omega}_{\bz\bz}\equiv \int e^{i\omega u}  \p_u C_{\bz\bz} du. 
\ee
Then from the large $r$ saddle point expansion of~(\ref{eq:saddle}), we have:
\begin{equation}
\begin{array}{lll}
N_{\bz\bz}^\omega&=& - \frac{ 1 }{2\pi } \hat{\e}_{\bz\bz}^{+} \omega a^{\mathrm{out}}_- (\omega {\hat x }),   \\
N_{\bz\bz}^{-\omega}&=& - \frac{ 1 }{2\pi } \hat{\e}_{\bz\bz}^{+} \omega a^{\mathrm{out}}_+ (\omega {\hat x })^\dagger,
\end{array}
\end{equation}
with $\omega>0$ in both cases.  We define $N^{(1)}_{\bz\bz}$ as:  
\begin{equation}
\begin{array}{ll}
N^{(1)}_{\bz\bz}&\equiv\int duuN_{\bz\bz}\\&=-\lim\limits_{\omega\rightarrow0}\frac{i}{2}(\partial_\omega N^\omega_{\bz\bz}+\partial_{-\omega} N^{-\omega}_{\bz\bz})\\
&=\frac{i}{4\pi} \hat{\e}_{\bz\bz}^{+} \lim\limits_{\omega\rightarrow0}(1+\omega\partial_\omega)[ a^{\mathrm{out}}_-(\omega\hat{x})-a^{\mathrm{out}}_+(\omega\hat{x})^\dagger].\\
\end{array}
\end{equation}
A mode expansion analogous to~(\ref{eq:czz}) can be defined for $D_{\bz\bz}$ on $\mathcal{I}^-$
\be
\label{eq:dzz}
D_{\bz\bz}=-\frac{i}{4\pi^2}\hat{\e}_{\bz\bz}^{+}\int_0^\infty d\omega_q[a^{\mathrm{in}}_-(\omega_q\hat{x})e^{-i\omega_qv}-a^{\mathrm{in}}_+(\omega_q\hat{x})^\dagger e^{i\omega_qv}],
\ee
from which we find
\begin{equation}
\begin{array}{lll}
M_{\bz\bz}^\omega&=& - \frac{ 1 }{2\pi } \hat{\e}_{\bz\bz}^{+} \omega a^{\mathrm{in}}_- (\omega {\hat x }),  \\
M_{\bz\bz}^{-\omega}&=& - \frac{ 1 }{2\pi } \hat{\e}_{\bz\bz}^{+} \omega a^{\mathrm{in}}_+ (\omega {\hat x })^\dagger,
\end{array}
\end{equation}
and
\begin{equation}
\begin{array}{ll}
M^{(1)}_{\bz\bz}&=-\lim\limits_{\omega\rightarrow0}\frac{i}{2}(\partial_\omega M^\omega_{\bz\bz}+\partial_{-\omega} M^{-\omega}_{\bz\bz})\\
&=\frac{i}{4\pi} \hat{\e}_{\bz\bz}^{+}\lim\limits_{\omega\rightarrow0}(1+\omega\partial_\omega)[ a^{\mathrm{in}}_-(\omega\hat{x})-a^{\mathrm{in}}_+(\omega\hat{x})^\dagger].\\
\end{array}
\end{equation}
We are interested in the  matrix element 
\begin{equation}
\label{eq:co}
\begin{array}{ll}
&\langle out |N^{(1)}_{\bz\bz}  \mathcal{S}+\mathcal{S}M^{(1)}_{\bz\bz} | in \rangle \\
  &=    \frac{ i}{4 \pi }\hat{\e}_{\bz\bz}^{+}\lim\limits_{\omega \to 0}  (1+\omega \p_\omega) 
\langle out  |  ( a^{\mathrm{out}}_- (\omega {\hat x })  - a^{\mathrm{out}}_+ (\omega {\hat x })^\dagger)\mathcal{S}+\mathcal{S}( a^{\mathrm{in}}_- (\omega {\hat x })  - a^{\mathrm{in}}_+ (\omega {\hat x })^\dagger)  | in \rangle\\
&=  \frac{ i  }{4\pi }\hat{\e}_{\bz\bz}^{+} \lim\limits_{\omega \to 0}  (1+\omega \p_\omega) 
\langle out  |  a^{\mathrm{out}}_- (\omega\hat{x})\cs-\cs a^{\mathrm{in}}_+ (\omega\hat{x})^\dagger  | in \rangle,\\  
\end{array}
\end{equation}
which is $\langle out | \mathcal{S}| in \rangle$ with soft graviton insertions.\footnote{ Here we assume that  $|in\> $  and $\<out|$ states contain no soft gravitons.}  Such insertions generically have Weinberg poles behaving as ${1 \over \omega}$.  However the prefactor $1+\omega\p_{\omega}$ projects out this pole, leaving   the subleading $\mathcal{O}(\omega^0)$ soft factor.  Equation~(\ref{eq:co}) and its hermitian conjugate are related to the 
$Q_{S}$ matrix element by 
\be
\label{eq:ot}
\begin{array}{lll}
&\langle out|  :Q_{S} \mathcal{S}: | in \rangle\\
&=  -\frac{1}{2}  \int d^2z \gamma^{z\bz}D_{z}^3 Y^{z}  \langle out |N^{(1)}_{\bz\bz}  \mathcal{S}+\mathcal{S}M^{(1)}_{\bz\bz} | in \rangle 
\\
&=-\frac{i}{8\pi}
\lim\limits_{\omega \to 0}  (1+\omega \p_\omega)  \int d^2z  \gamma^{z\bz}D_{z}^3 Y^{z} 
\hat{\e}_{\bz\bz}^{+} \langle out  |   a^{\mathrm{out}}_- (\omega\hat{x})\cs-\cs a^{\mathrm{in}}_+ (\omega\hat{x})^\dagger  | in \rangle .
\end{array}
 \ee
Given the asymptotic behavior~(\ref{eq:asym}) near $i^0$, the boundary relation $N_{\bz\bz}|_{\ci^+_-}=-M_{\bz\bz}|_{\ci^-_+}$ establishes a correspondence between the in and out modes, such that the contributions to the matrix element~(\ref{eq:ot}) from the $a^{\mathrm{out}}_- (\omega\hat{x})$ and $-a^{\mathrm{in}}_+ (\omega\hat{x})^\dagger$ insertions are equal.

\section{From soft theorem to Virasoro symmetry}

In this section we begin by assuming the subleading-soft relation\footnote{A single soft graviton insertion has the $\omega$ expansion 
\begin{equation*}
\langle z_{n+1},z_{n+2},...|a_- (q)\cs|  z_{1},z_{2},...\rangle =\left( S^{(0)-} +S^{(1)-}\right) \langle  z_{n+1},z_{n+2},...  | \cs | z_{1},z_{2},... \rangle + \co(\omega).
\end{equation*}
}
 \be
\label{eq:n}
\begin{array}{ll}
&\lim\limits_{\omega \to 0}  (1+\omega \p_\omega)\langle z_{n+1},z_{n+2},...|a_- (q)\cs|  z_{1},z_{2},...\rangle  = S^{(1)-}
 \langle  z_{n+1},z_{n+2},...  | \cs | z_{1},z_{2},... \rangle, 
 \end{array}
\ee
 with
 \bea \label{sone}S^{(1)-}&=&-i\sum_k\frac{p_{k\mu}\epsilon^{-\mu \nu} q^\lambda J_{k\lambda \nu}}{p_k\cdot q}.
 \eea
Here 
$
J_{k\lambda\nu} \equiv L_{k\lambda\nu} +S_{k\lambda \nu} 
$
is the total ingoing orbital+spin angular momentum of the k$^{th}$ particle which obeys the global conservation law  $\sum J_{k\lambda\nu}=0$.  We note the $(1+\omega \p_\omega)$ prefactor on the left hand side projects out the would-be Weinberg pole accompanying a soft insertion.  For notational brevity we consider the contribution for negative polarization: the general formula has an $S^{(1)}$ with a general polarization tensor replacing (\ref{sone}). We will show that (\ref{eq:n}) implies the Ward identity (\ref{conj}), which in turn is equivalent to infinitesimal \x-invariance of the $\cs$-matrix. 
Although the relation (\ref{eq:n}) potentially has wider validity, the only case in which it is known to be a theorem is tree-level gravitons \cite{fc}.  Hence  only for this case do we claim the results of this section imply a Virasoro symmetry.  

Gauge invariance provides an important check on this formula.  Amplitudes must vanish for pure gauge gravitons with polarizations 
\be \e_\Lambda^{\mu\nu}=q^\mu\Lambda^\nu + q^\nu\Lambda^\mu \ee
for any $\Lambda$. Inserting this into (\ref{sone}) we find
\be iS^{(1)}(\e_\Lambda) = q^\mu \Lambda^\nu \sum_k J_{k\mu\nu}+  \sum_k\frac{p_{k}\cdot \Lambda  q^\mu q^\nu J_{k\mu\nu}}{p_k\cdot q}.\ee
The first terms vanishes by global angular momentum conservation, while the second vanishes by antisymmetry of $J_{k\mu\nu}$. This is very similar to the gauge invariance of the Weinberg pole, which vanishes due to global energy-momentum conservation or equivalently  translational symmetry.  The Weinberg soft theorem  implies that this global translational symmetry is promoted to a local supertranslational symmetry on the sphere \cite{hms}, because there is one symmetry for every angle $\vec q$. In this section we will see a parallel story for rotational invariance: the soft relation (\ref{eq:n}) implies that rotations are promoted to a local superrotational - equivalently Virasoro - symmetry on the sphere.

The first step is to write the hard particle momenta  $p_k$, the soft graviton momentum $q$ and chosen polarization $\e^{-\mu\nu}=\e^{-\mu}\e^{-\nu}$ in terms of the  points $z_k$ and $z$ at which they  arrive on the 
on the asymptotic $S^2$ and their energies $E_k,\omega$
\be
\begin{array}{ll}
p_k^\mu&=\frac{E_k}{1+z_k\bar{z}_k}\left(1+z_k\bar{z}_k,\bz_k+z_k,i(\bar{z}_k-z_k),1-z_k\bar{z}_k\right),\\
q^\mu&=\frac{\omega}{1+z\bar{z}}\left(1+z\bar{z},\bar{z}+z,i(\bar{z}-z),1-z\bar{z}\right),\\
\epsilon^{-\mu}&=\frac{1}{\sqrt{2}}(z,1,i,-z).
\end{array}
\ee
 One then finds for the orbital terms 
\be
S^{(1)-}=\sum_k\left(\frac{E_k(z-z_k)(1+z\bar{z}_k)}{(\bz_k-\bz)(1+z_k\bar{z}_k)}\partial_{E_k}+
\frac{ (z-z_k)^2}{(\bz_k-\bz)}\partial_{z_k}\right).
     \ee 
The spin term will be added in below.  This expression obeys 
\be
\gamma^{z\bar{z}}D_{z}^3(\hat{\e}_{\bz\bz}^{+}S^{(1)-})=-2\pi \sum_k\bigl( D_{z}\delta^{(2)}(z-z_k)E_k\partial_{E_k}
+2\delta^{(2)}(z-z_k)\partial_{z_k}\bigr).
\ee
Multiplying both sides of (\ref{eq:n}) by $D_z^3Y^z \hat{\e}^{+z}_{~~\bz}$ and integrating over the soft graviton angle $z$ gives
\be\label{conjd}
\begin{array}{ll}
&\<z_{n+1},z_{n+2},... |:Q_S \mathcal{S}:|z_1,z_2,...\>=\\
&~~~~~~~~~~~~~~~~~ - i\sum\limits_k \left( Y^z(z_k)\p_{z_k}-\frac{E_k}{2} D_zY^z(z_k) \p_{E_k}\right) \<z_{n+1},z_{n+2},...|\mathcal{S} |z_1,z_2,...\>,
\end{array}\ee
which is exactly the Ward identity (\ref{conj}) arising from an asymptotic Virasoro symmetry, minus the so-far-omitted spin terms. 

The spin contribution comes from evaluating:
\be
S^{(1)-}_{S}=-i\sum_k\frac{p_{k\lambda } \epsilon^{-\lambda\nu} q^\mu S_{k\mu\nu}}{p_k\cdot q}.
\ee
In terms of the helicity $h$ defined by
\be   h p_\mu  = -\frac{1}{2}\e_{\mu\nu\lambda \rho  } S^{\nu\lambda} p^\rho, \ee
one finds
\be
\label{eq:s1}
S^{(1)-}_{S}=\sum_k \frac{(z-z_k)(1+z\bar{z}_k)}{(\bz-\bz_k)(1+z_k\bar{z}_k)}h_k,\ee
 while the third derivative obeys
\be
\gamma^{z\bar{z}}D_{z}^3(\hat{\e}_{\bz\bz}^{+}S^{(1)-}_{S})=2\pi \sum_k h_k D_{z}\delta^{(2)}(z-z_k).
\ee
Hence the spin contribution  for the helicity states  corrects (\ref{conjd}) to: 
\be\label{dki}
\begin{array}{ll}
\<z_{n+1},z_{n+2},... |:Q_S \mathcal{S}:|z_1,z_2,...\>&= -i \sum\limits_k \left( Y^z(z_k)\p_{z_k}-\frac{E_k}{2} D_zY^z(z_k) \p_{E_k}+\frac{ h_k }{2} D_zY^z(z_k)\right) \\ 
&~~~~~~~~~~~~~~~~~~~~~~~~~~~\<z_{n+1},z_{n+2},...|\mathcal{S} |z_1,z_2,...\>,
\end{array}
\ee
in agreement with the spin-corrected version of (\ref{conj}).

In conclusion the soft relation (\ref{eq:n}), whenever valid,  implies a Virasoro symmetry of the quantum gravity $\cs$-matrix.

\section*{Acknowledgements}
We are  grateful to Z. Bern, J. Bourjaily, F. Cachazo, S. Caron-Huot, C. Cordova, T. Dumitrescu, T. He,  J. Maldacena, P. Mitra and M. Schwartz for useful conversations.   This work was supported in part by DOE grant DE-FG02-91ER40654  and the Fundamental Laws Initiative at Harvard.

\end{document}